\documentclass[a4paper]{spie}  
\usepackage{graphicx}
\usepackage{amsmath,amssymb,amsfonts}       

\usepackage{color}

\newcommand{\average}[1]{\left \langle {#1} \right \rangle}

\newcommand{\T}{\mathsf{T}} 
\newcommand{\I}{\mathbf{I}} 
\newcommand{\Hkal}{\mathcal{H}_{\infty}} 
\newcommand{\Lkal}{\mathcal{L}_{\infty}} 
\newcommand{\K}{\mathcal{K}_{\infty}} 
\newcommand{\z}{{z}} 

\newcommand{\Proj}{\mathbf{H}} 
\newcommand{\Rec}{\mathbf{R}} 
\newcommand{\Emv}{\mathbf{E}} 
\newcommand{\CovMat}{\boldsymbol{\Sigma}} 

\newcommand{\D}{\mathbf{G}} 

\newcommand{\tr}{\operatorname{trace}} 

\newcommand{\dint}{\mathrm{d}} 

\newcommand{\Ad}{\mathcal{A}_\mathsf{tur}} 
\newcommand{\Bd}{\mathcal{B}_\mathsf{tur}} 
\newcommand{\Cd}{\mathcal{C}_\mathsf{tur}} 
\newcommand{\Dd}{\mathcal{D}_\mathsf{tur}} 

\newcommand{\alphavec}{{\boldsymbol{\alpha}}} 
\newcommand{\betavec}{{\boldsymbol{\beta}}} 
\newcommand{\rhovec}{\boldsymbol{\rho}}
\newcommand{\etavec}{{\boldsymbol{\eta}}} 
\newcommand{\phivec}{{\boldsymbol{\psi}}} 
\newcommand{\varphivec}{{\boldsymbol{\varphi}}} 
 
\newcommand{\svec}{{\mathbf{s}}}

\newcommand{\qed}{\nobreak \ifvmode \relax \else
      \ifdim\lastskip<1.5em \hskip-\lastskip
      \hskip1.5em plus0em minus0.5em \fi \nobreak
      \vrule height0.75em width0.5em depth0.25em\fi}
      
\title{Natural guide-star processing for wide-field laser-assisted AO systems} 
 
\author{Carlos M. Correia\supit{a}, Benoit Neichel\supit{a},
  Jean-Marc Conan\supit{b}, Cyril Petit\supit{b},  Jean-François
  Sauvage\supit{a,b}, Thierry Fusco\supit{a,b}, Joel
  D. R. Vernet\supit{c} and Niranjan Thatte\supit{d}
\skiplinehalf
\supit{a}Aix Marseille Universit\'e, CNRS, LAM  (Laboratoire d'Astrophysique de Marseille) UMR 7326, 13388 Marseille, France; \\
\supit{b} ONERA (Office National d'Etudes et de Recherches A\'erospatiales), B.P.72, F-92322 Ch\^atillon, France; \\
\supit{c} European Southern Observatory, Karl-Schwarzschild-Strasse 2, D-85748 Garching b. Munchen, Germany \\
\supit{d} Dept. of Astrophysics, University of Oxford, Keble Road, Oxford, OX1 3RH, United Kingdom
}

\authorinfo{Further author information: carlos.correia@lam.fr}

\begin{document} 
  \maketitle 

\begin{abstract}

Sky-coverage in laser-assisted AO observations largely depends on the
system's capability to guide on the faintest natural guide-stars
possible. Here we give an up-to-date status of our natural guide-star
processing tailored to the European-ELT's visible and near-infrared (0.47
to 2.45 $\mu m$) integral field spectrograph -- Harmoni.

We tour the processing of both the isoplanatic and anisoplanatic tilt
modes using the spatio-angular approach whereby the wavefront is
estimated directly in the pupil plane avoiding a cumbersome explicit
layered estimation on the 35-layer profiles we're currently using.

Taking the case of Harmoni, we cover the choice of wave-front sensors,
the number and field location of guide-stars, the optimised algorithms
to beat down angular anisoplanatism and the performance obtained with
different temporal controllers under split high-order/low-order
tomography or joint tomography. We consider both atmospheric and far
greater telescope wind buffeting disturbances. In addition we provide
the sky-coverage estimates thus obtained. 
\end{abstract}



\section{Introduction}In this work we layout the NGS modes processing for laser tomography AO
systems. We start by addressing the common tilt mode (isoplanatic)
considering different control strategies from the ESO-suggested double
stage, double integrators with lead filters and
Linear-Quadratic-Gaussian (LQG) controllers under two distinct
operating scenarios: in stand-alone mode or in tandem with the
telescope's image stabilisation controller averaging tip/tilt
signals from three NGS off-axis.

We then move on to handle the anisoplanatic tilt. Three models are presented: \\
\textit{i)} tilt-tomography using a
combination of tilt and high-altitude quadratic modes that produce
pure tilt through cone projected ray-tracing through the wave-front
profiles \cite{correia13}; \\
\textit{ii)} a spatio-angular MMSE tilt estimation
anywhere in the field that is more general. \\
\textit{iii)} the
(straightforward) generalisation to dynamic controllers using
near-Markovian time-progression models from \cite{correia15}. 

These controllers will be used for the HARMONI NGS modes
\cite{thatte16, neichel16, sauvage16}.

\section{ Design of HARMONI (E-IFU) tilt (an)isoplanatism controllers}

\subsection{AO loop transfer functions}
We will model the AO loop as independent transfer-function
mode-per-mode ignoring any eventual cross-spectrum recurring to the
customarily-used Laplace-transformed transforms and variables $s=2\pi i \nu$ with $\nu \in
[0:1/(2T_s)]$ the temporal frequency vector. For the cases where the discrete Z-transform needs be used we assume $z = e^{2i \pi \nu}$.

The open-loop transfer function is assembled as follows \cite{roddier99}
\begin{equation}
h_{ol} (s)= h_{WFS} h_{dac} h_{lag} h_{ctr}(h_{low}h_{m5}+ (1-h_{low})h_{m4})
\end{equation} 
where the tilt correction is provided jointly by the E-ELT's M4 and M5
mirrors whose transfer functions are \cite{sedghi16}
\begin{equation}
h_{m4} (s) = \frac{w^2}{s^2 + 2\xi w s + w^2}, \xi = 0.35 , w = 2\pi\times500 [rad/s]
\end{equation}
and 
\begin{equation}
h_{m5} (s) = \frac{w}{s + w}, w = 2\pi\times10 [rad/s]
\end{equation}
We have opted for low temporally filtering the off-load to M5 using a
low-pass filter $h_{low}$, which we take to be a order-1 filter
\cite{correia12}.

In case the controller  $h_{ctr}$ is the double-integrator+lead filter \cite{correia13}
\begin{equation}
h_{ctr} (s) = g \left( \frac{1}{1-e^{-2\pi \nu T_s}}\right)^2
\left(\frac{1+2\pi\nu T_l}{1-2\pi\nu T_l}\right), T_l = \frac{1}{\sqrt{2\pi \nu_0 \sqrt{a}}}
\end{equation}
with $\{\nu_o, T_l, a\}$ the lead filter parameters to be optimised \cite{correia12b}. 

For the LQG case the synthesis is done in discrete-time  
leading to \cite{correia10a}
\begin{equation}
h_{ctr} (\z) =  - \left[ \I + \K \Lambda_{p} \left(\z^{-d} \Lkal
      \mathcal{D} + \Bd \right)\right]^{-1} \K
            \Lambda_{p} \Ad \Hkal  ,
\end{equation}
 where 
  \begin{equation}
    \Lambda_{p} = \left(\I -  \z^{-1}\Ad \left( \I - \Hkal \Cd
              \right)\right)^{-1} .
  \end{equation}
with $\K \triangleq \Ad \Hkal $ the solution of the discrete algebraic
Riccati equation\cite{andersonmoore_optimalcontrolLQG05} and $\{\Ad,
\Bd, \Cd, \Dd\}$ the compound matrices of discrete-time state-space
model and $d$ the delay in integer multiples of the sampling interval
$T_s$.

One other option is considered below, namely ESO's double-stage
controller [9911-39], this conference \cite{sedghi16}.

\subsection{Measurement model with time-averaged variables }\label{sec:measurement-model}


We assume a straight Zernike-to-slopes static model (i.e. no temporal
averaging as done elsewhere \cite{correia13}); the modal matrix $\D$ translates modal coefficients of TT modes into average slopes over the illuminated sub-region of
each sub-aperture with
\begin{align}\label{eq:GammaTT}
  \D & \triangleq \left( 
       \begin{array}{cc}
         \gamma_\textsf{T} & 0\\
         0&  \gamma_\textsf{T}  
       \end{array}%
            \right) 
\end{align}
where $ \gamma_\textsf{T}  = 2$ for full-aperture TT-WFS and can easily
be generalised when more sub-apertures are present in the SH-WFS.


\subsection{Noise model}
We have simulated tilt-removed random draws from the residual
tomographic PSDs
across the field simulated with Fast-F \cite{neichel09} for multiple
wavelengths. We consider SH-WFS with 1, 2, 4, 10 and 20 sub-apertures (linear).

\begin{figure}[htpb]
	\begin{center}
            \hspace{-3.6cm}\includegraphics[width=1.2\textwidth]{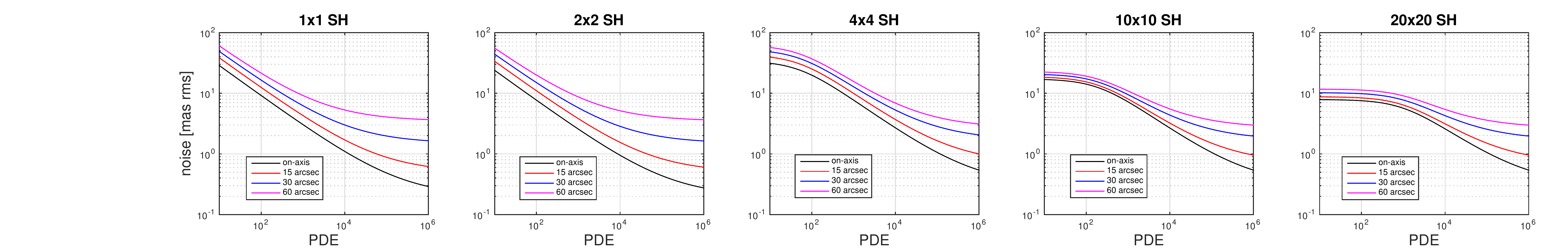}
	\end{center}
	\caption[]
	{\label{fig:noiseCurves}
   Noise curves comparing the model in this section to FAST-F computed
   noise levels as a function of the PDE (photon-detection equivalent)
   for full-pupil SH (1x1) to 20x20 SH operating in H-band. Saturation occurs on low-flux
   with increasing number of sub-apertures. Aliasing is seen on the
   high-end flux cases due to aliasing. }
\end{figure}

Figure \ref{fig:noiseCurves} shows the noise curves fitted to the
simulated ones for all the LO-WFS cases considered -- with
photon-noise only for the time being. These curves are in good agreement with standard expressions (in angle
rms units)
\begin{equation}\label{eq:etab}
  \sigma_\eta = \frac{\theta_b}{\textsf{SNR}},   [rad rms]
\end{equation}
where $\theta_b$ is the effective spot size of the sub-aperture and SNR is the signal-to-noise ratio of a
single sub-aperture, when the field-dependent Strehl-ratio (SR) and
PSF FWHM across the field are taken
into account. However, these curves have a major advantage: non-linear
effects like saturation at low flux levels and aliasing causing the
roll-off at high-fluxes are taken into account. The behaviour far from
these extremes is, as expected, $\propto n_{ph}^{-1/2}$

\section{Isoplanatic tilt correction}\label{sec:isoTilt}

\subsection{Isoplanatic input disturbances}
\begin{figure}[htpb]
	\begin{center}
            \includegraphics[width=0.58\textwidth]{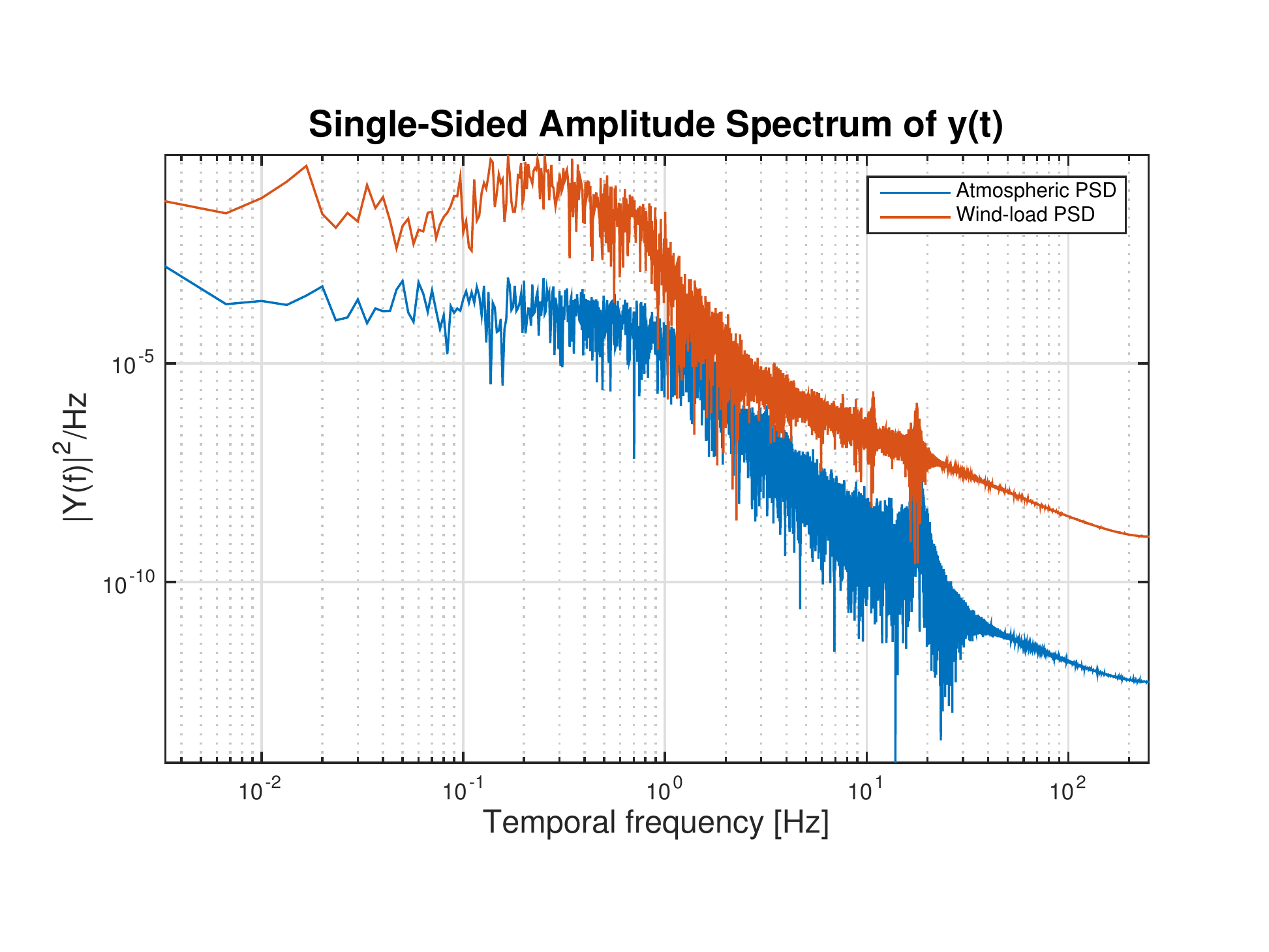}
	\end{center}
	\caption[]
	{\label{fig:psdTTAtmWindLoad}
   Power-spectral densities for the atmosphere-induced and
   wind-induced image jitter. The latter was simulated with wind face on aligned with the main
axis are depicted (ESO internal report). For the meadian seeing
conditions simulated (see below Fig. 9 for the Cn2 profiles) we get
$\sim$ 20 mas rms and $\sim$270 mas rms input disturbances
respectively.}
\end{figure}

We consider stationary and non-stationary sources of tilt. The former
are depicted in Fig. \ref{fig:psdTTAtmWindLoad} where the temporal PSD
of  both the turbulent
tilt modes causing image jitter and wind-buffeting on the telescope
structure cause image motion.

\begin{figure}[htpb]
	\begin{center}
            \includegraphics[width=0.58\textwidth]{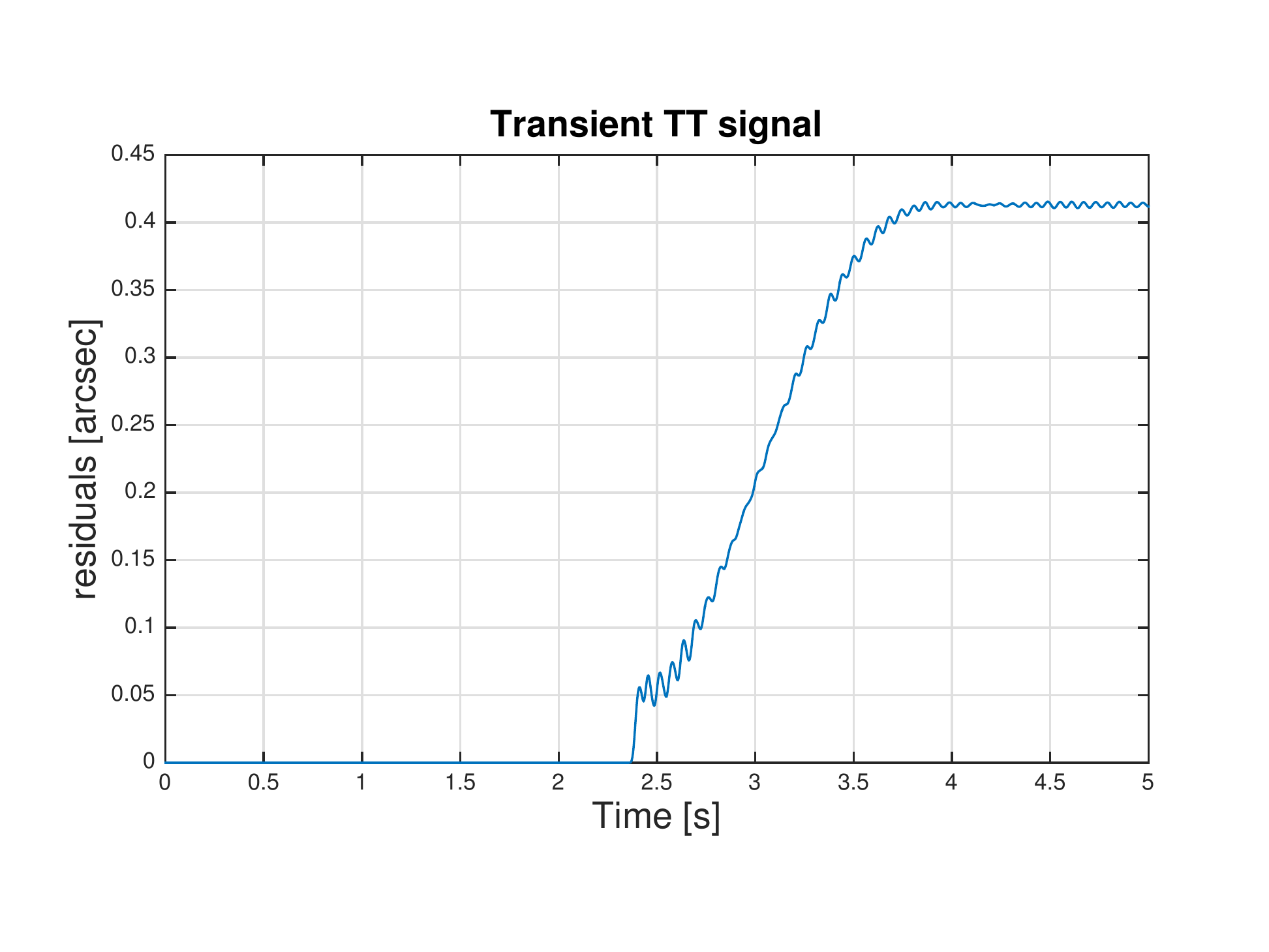}
	\end{center}
	\caption[]
	{\label{fig:transient}
   Transient signal affecting M4 every 5 minutes. }
\end{figure}

We further consider the non-stationary tilt from M1-3 off-loading to M4
causing a transient signal of about 400 mas rms roughly every 5 min (baseline)
represented in Fig \ref{fig:transient}.
\subsection{Handover scenarios}

Isoplanatic tilt will be corrected at the telescope level by up to
three off-axis tilt sensors guiding on faint stars not vignetting the
instrument's field. 

We are currently evaluating two scenarios
\begin{enumerate}
\item \textbf{sequential handover}, where the instrument tilt signal alone
  drives M4/M5 after a transitioning period from the telescope over to
  the instrument
\item \textbf{cascaded handover}, where the off-axis tilt signals from the telescope and
  the on-axis (or closer in) instrument tilt are mingled to drive M4/M5 
\end{enumerate}

Although we do not cover the transitioning period in this work, we
note the approach outlined in Raynaud et al \cite{raynaud16}
\subsection{Sequential handover}
We consider four control strategies broadly cited in the literature \cite{correia12a,correia11e,correia12b}
\begin{enumerate}
\item double-stage composed of two single integrators and off-load loops 
\item optimised gain double integrator with lead filter
\item LQG with i) AR1 and ii) AR2 temporal evolution models
\end{enumerate}
The last three are coupled to a order-1 temporal filter with a 10\,Hz
cut-off frequency (subject to optimisation not done here although possible\cite{correia12a}) that low-pass filters tilt signals to M5 and assigns
its complement to M4.

\begin{figure}[htpb]
	\begin{center}
            \includegraphics[width=0.8\textwidth]{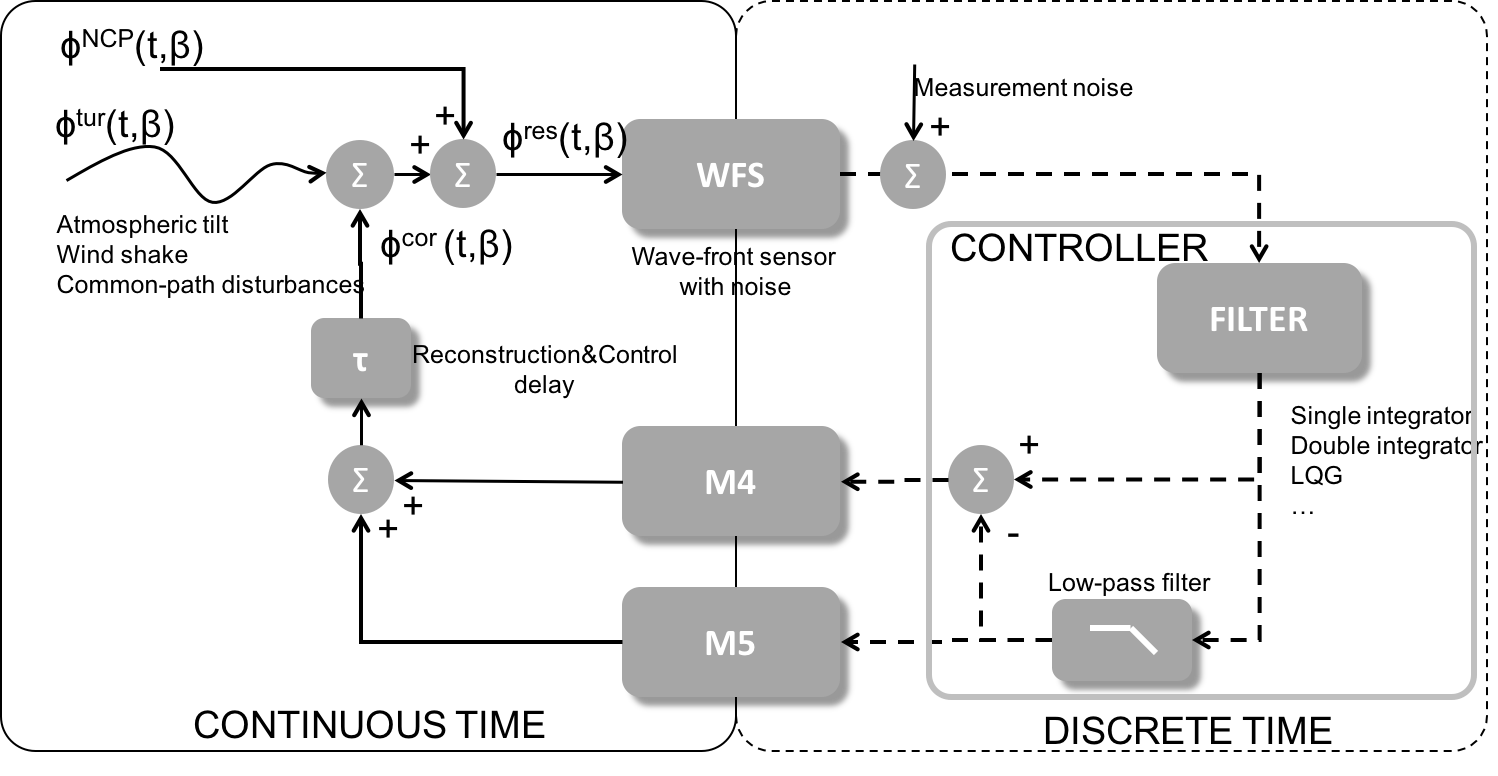}
	\end{center}
	\caption[]
	{\label{fig:Hdiagram}
     General block diagram for TT control. The low-pass filter is subject to optimisation in terms of its cut-off frequency and model order.}
\end{figure}

Figure \ref{fig:Hdiagram} provides a block-diagram for
integrator-based and LQG controllers.

\begin{figure}[htpb]
	\begin{center}
            \includegraphics[width=0.58\textwidth]{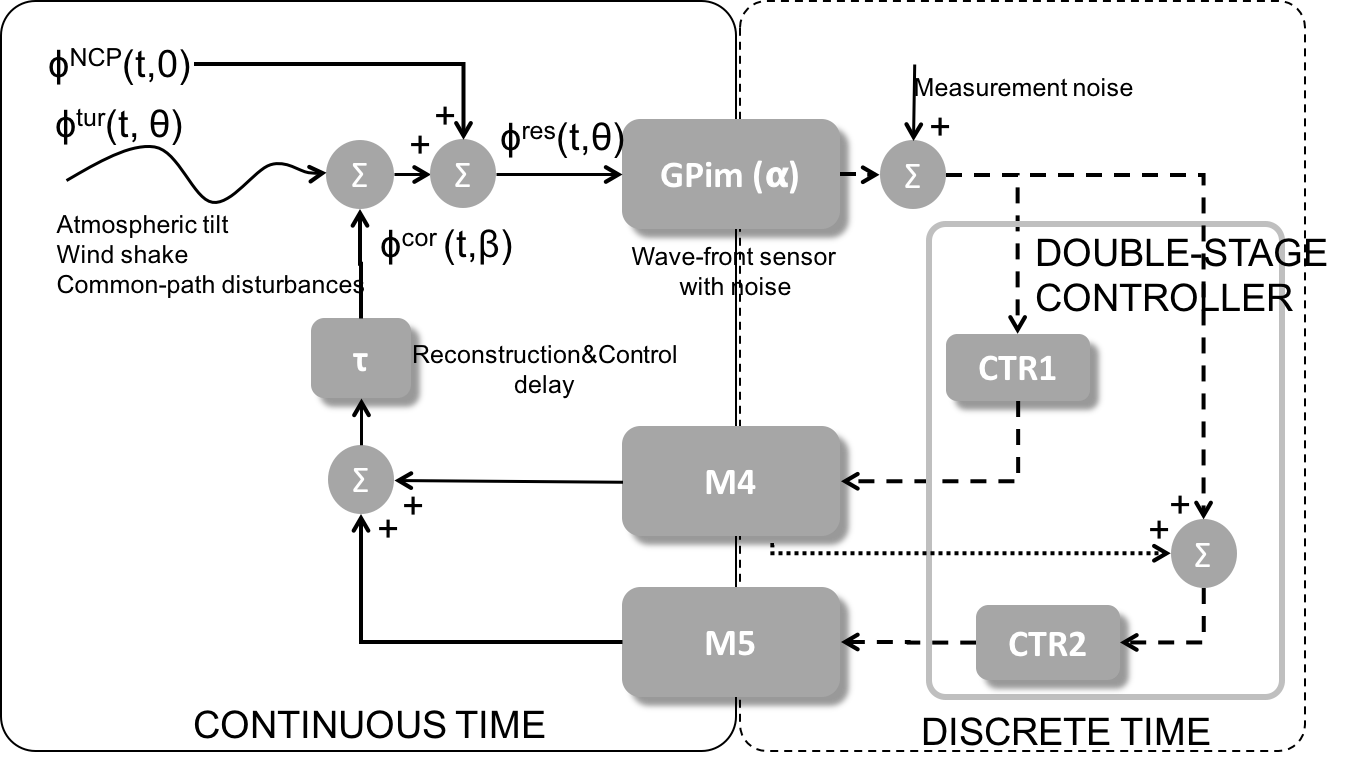}
	\end{center}
	\caption[]
	{\label{fig:doubleStageDiagram}
     Double-stage block diagram for TT control where CRT1 and CRT2 are
     integrators of the kind $g/({1 - e^{-T_s s} })$. The electric
     signal off from M4 consists in positioning sensors outputs.}
\end{figure}

For the double-stage controller, a different off-loading strategy is
obtained with two single integrators as is shown in
Fig. \ref{fig:doubleStageDiagram}, but with an overall low-frequency
rejection typical of a double-integrator i.e. 20dB/decade.

\subsubsection{Atmospheric and wind-induced tilt handling}

The rejection transfer function is gathered as
\begin{equation}\label{eq:RTF}
H_{RTF} = \left| \frac{1}{1+h_{ol}}\right|^2
\end{equation}
and 
\begin{equation}\label{eq:NTF}
H_{NTF} = \left| \frac{h_{sys}}{1+h_{ol}}\right|^2
\end{equation}
where $h_{sys} = h_{ol}/h_{wfs}$.
 
\begin{figure}[htpb]
	\begin{center}
            \includegraphics[width=0.58\textwidth]{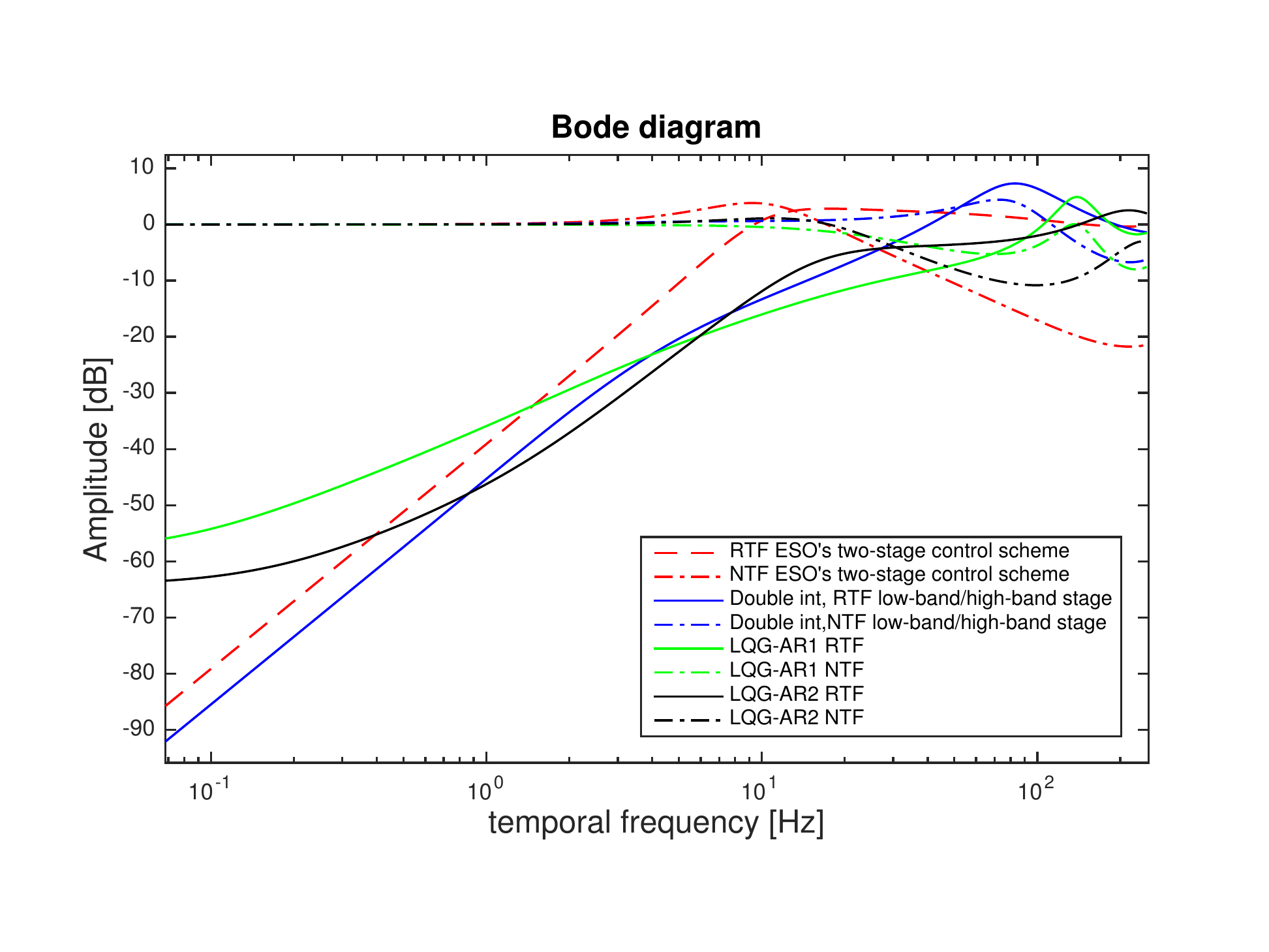}
	\end{center}
	\caption[]
	{\label{fig:controllerRtfNtf}
     Controller's rejection and noise transfer functions.}
\end{figure}

Figure \ref{fig:controllerRtfNtf} summarises the results thus far,
depicting the rejection and noise transfer functions. In Table
\ref{tab:comparisonControllers} numerical results are provided.

\begin{table}[htpb]
  \caption[] {\label{tab:comparisonControllers}
    Residuals in nm rms, noise propagation coefficients and stability margins. }
  \begin{center}
    \begin{tabular}{c c c c c c}
      \hline \hline
      Controller & Residual & Noise-propagation
                                      & gain margin &
                                                                 phase
                                                                 margin\\ 
      & mas rms & coefficient & & \\\hline
      Double stage & 2.6 & 0.17 & 16dB@73Hz & 44.5$^o$@13.1Hz \\
      Double-integrator+lead & 1.57 & 0.99 & 5.7dB@101Hz &  42.7$^o$@50.9Hz\\
      LQG + AR1 & 1.81 & 0.46& 7.5dB@145Hz  & 29.9$^o$@64.4Hz \\
      LQG + AR2 & 1.15 & 0.31& 12dB@215Hz & 149$^o$@30.2Hz \\
      \hline \hline
    \end{tabular}
  \end{center}
\end{table} 

\subsubsection{Transient handling}

\begin{figure}[htpb]
	\begin{center}
          \includegraphics[width=0.5\textwidth]{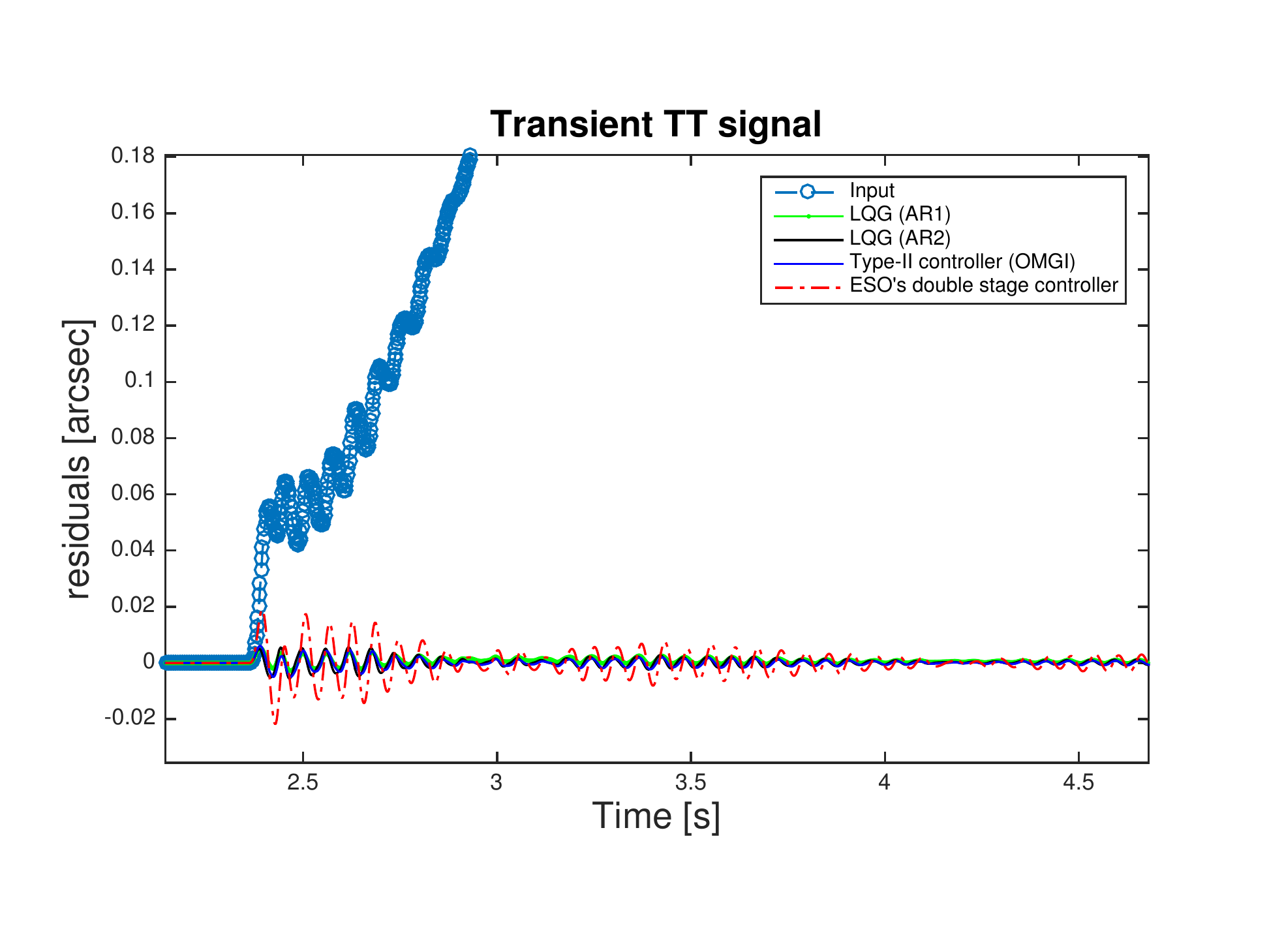}\includegraphics[width=0.5\textwidth]{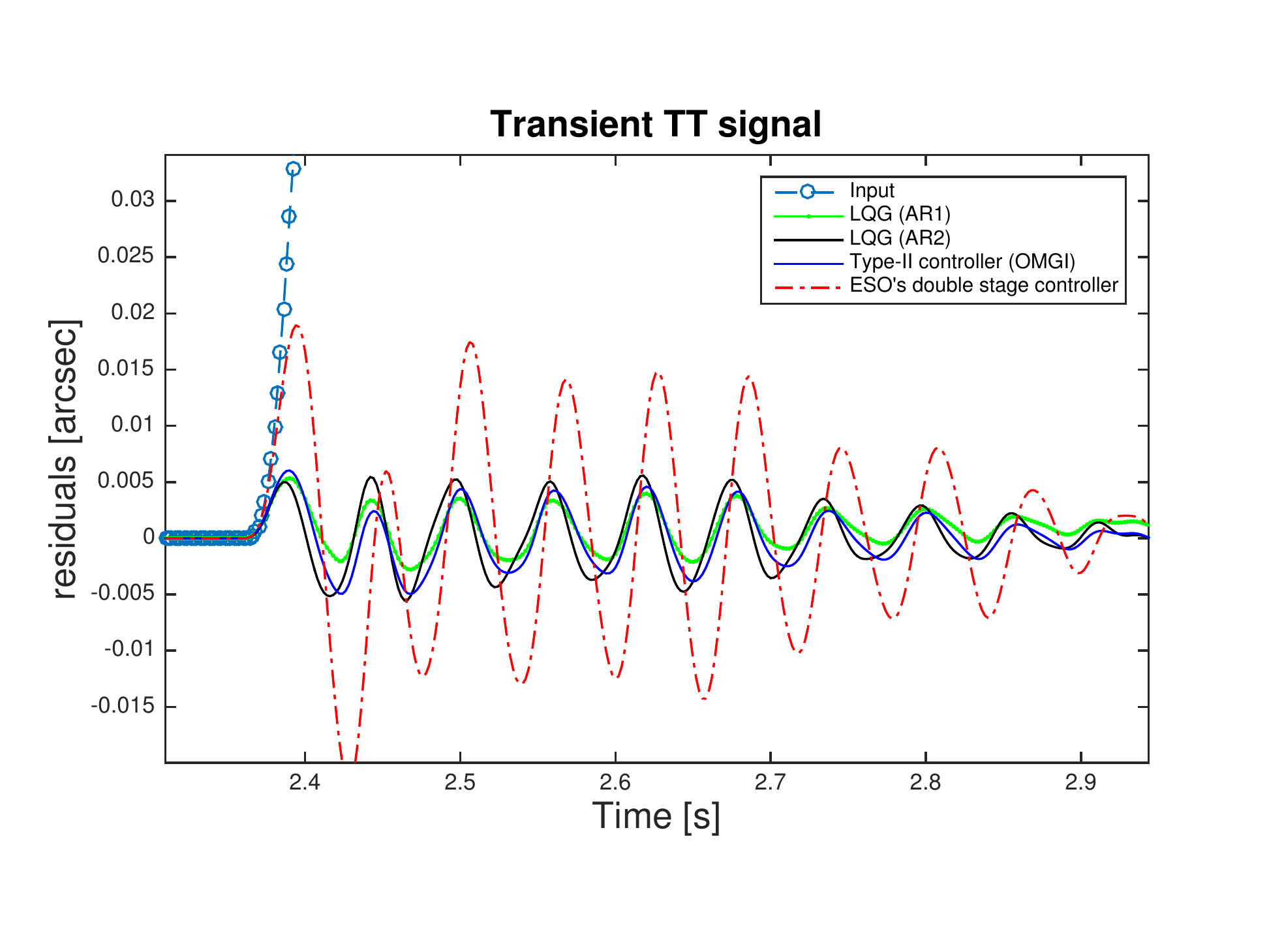}
	\end{center}
	\caption[]
	{\label{fig:transientHandling}
     Transient handling by the different controllers. The
     double-integrator+lead and the LQGs all keep the residual within
     $\pm$ 6 mas rms whereas the double-stage, on account of its
     limited closed-loop bandwidth achieves an intolerable 18 mas peak
     value.}
\end{figure}

When dealing with non-stationary signals as the transient caused by
off-loading M1-3-accumulated errors over to M4, transfer
function analysis can no longer be applied. We thus run the transient
through the controllers using time-domain simulations -- 
Fig. \ref{fig:transientHandling}.  Note all the controllers but the
double-stage keep the residual to within $\sim$6\,mas whereas the latter, on account of its limited temporal bandwidth presents a spike of roughly 3 times as much, i.e. $\sim$18\,mas.

\subsubsection{Vibrations handling}
Although we have processed various input signals with vibration peaks
following \cite{correia12} we defer to a later stage the full results
when spectra from the E-ELT becomes available \cite{sedghi16}.

\subsection{Cascaded handover}
In cascaded handover, M4\&M5 are driven jointly by the telescope and
instrument's tilt signals. 

\begin{figure}[htpb]
	\begin{center}
            \includegraphics[width=0.32\textwidth]{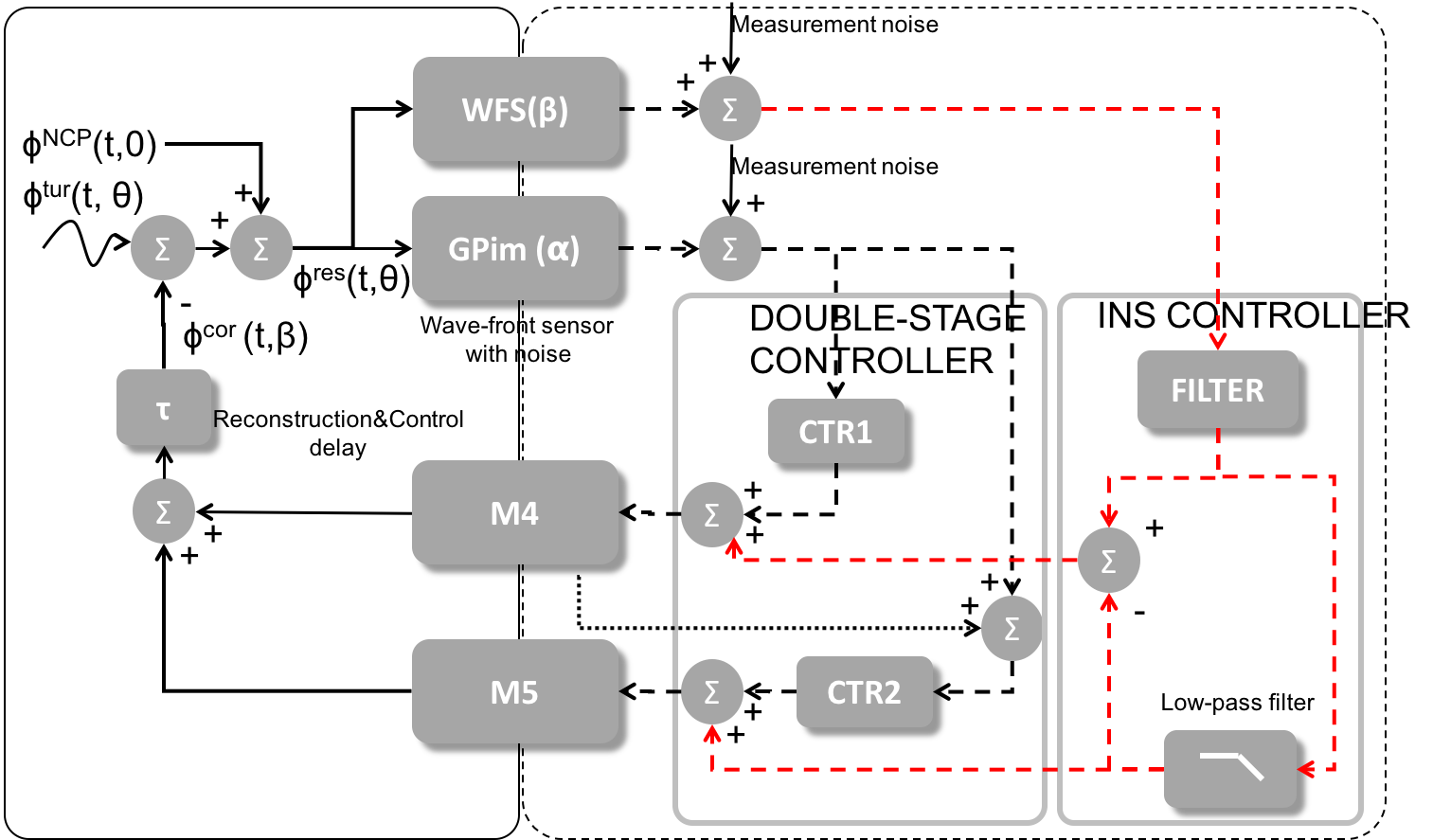}\includegraphics[width=0.29\textwidth]{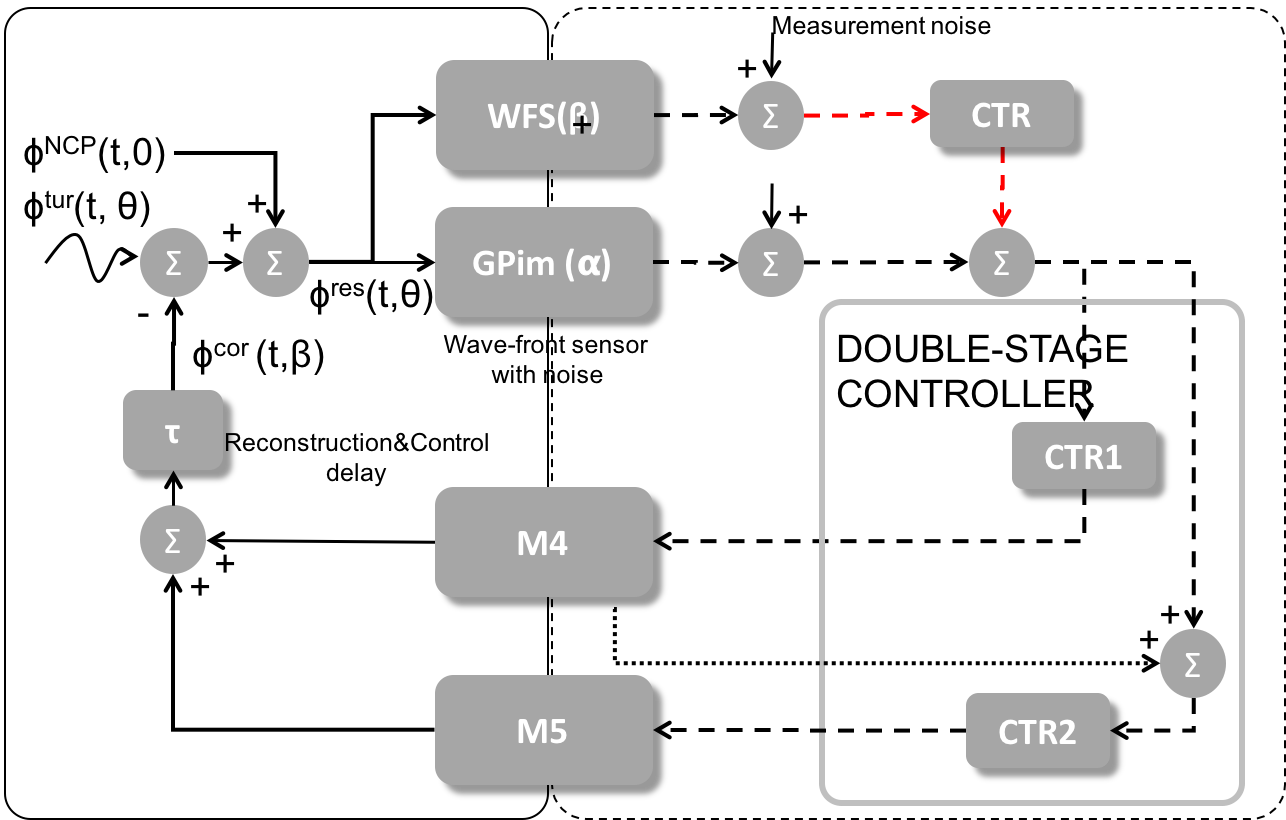} \includegraphics[width=0.28\textwidth]{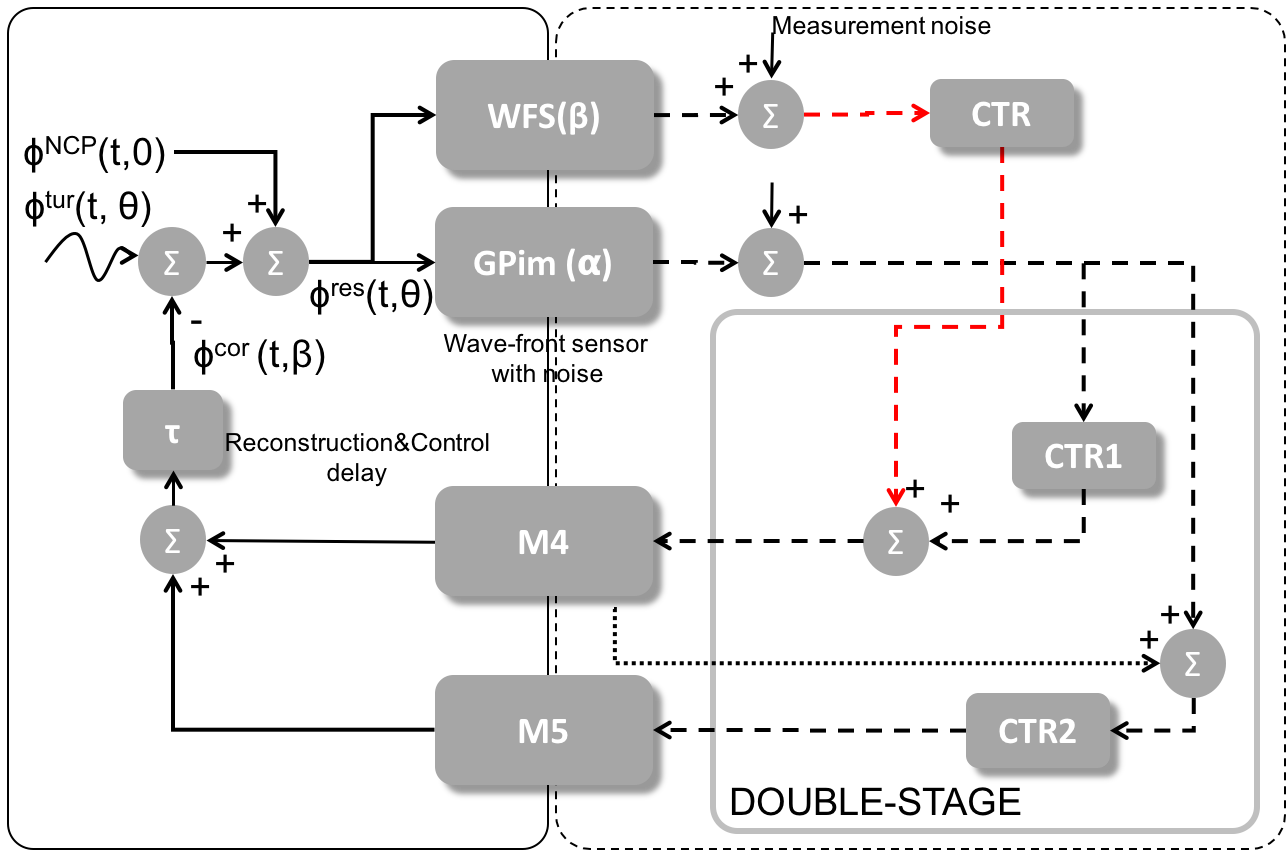}

\includegraphics[width=0.6\textwidth]{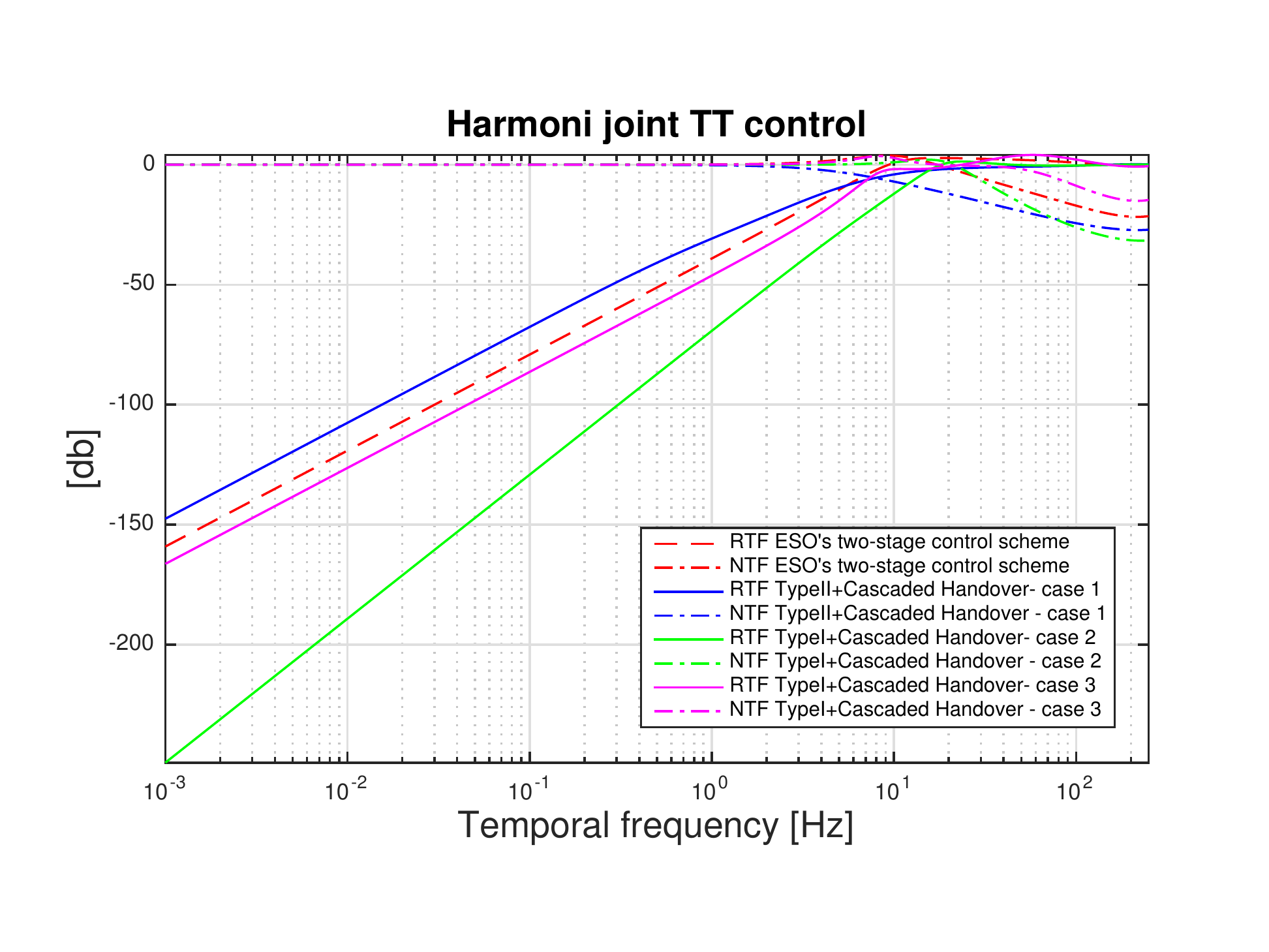}
	\end{center}
	\caption[]
	{\label{fig:cascadedHandover}
      (Top:) Three options tested for the cascaded handover
      scenario. (Bottom:) the corresponding rejection transfer
      functions when the instrument tilt controller is an
      optimised-gain double-integrator+lead-filter (blue-left, case
      1), whereas cases 2 and 3 (respectively green and magenta) used
      a single optimised-gain integrator. All cases with $\gamma$ =
      1/2. Red-dashed curves for the double-integrator in stand-alone
      mode for comparison.}
\end{figure}

We foresee three options
\begin{enumerate}
\item LOWFS signals are processed and filtered with a custom-filter
  and dispatched directly to M4 and M5 following the
  low-pass/high-pass strategy presented above. The overall open-loop transfer
  function becomes
\begin{equation}
h_{ol} = h_{WFS} h_{dac} h_{lag} \left\{ 
\begin{array}{c}
\left[ \gamma h_{C4} + (1-\gamma)h_{CTR} (1-h_{low})\right]h_{m4} + \\
\left[-\gamma + \left[-\gamma h_{C4} + (1-\gamma)h_{CTR}
      (1-h_{low})\right]h_{m4}\right]h_{C5}h_{m5} + \\
 (1-\gamma)h_{CTR}h_{low}h_{m5} 
\end{array}
\right\}
\end{equation}
where $\gamma \in [0\cdots 1]$ is a user-defined parameter that
balances the relative weight on the instrument LOWFS with respect to
the telescope off-axis tilt signals
\item instrument tilt is added in series to telescope tilt and later
  filtered by the double-stage
\begin{equation}
h_{ol} = h_{WFS} \left[ -\gamma + (1-\gamma)h_{CTR} \right] h_{dac}
h_{lag} \left( h_{C4}h_{m4} + h_{C5}h_{m5} +h_{C4}h_{m4} h_{C5}h_{m5} \right)
\end{equation}
\item instrument tilt is affected solely to M4, the double stage
  taking charge of off-loading it to M5 as is goes
\begin{equation}
h_{ol} = h_{WFS} h_{lag} 
\left[ -\gamma + (1-\gamma)h_{CTR} \right] h_{dac}
h_{lag} \left( h_{C4}h_{m4} + h_{C5}h_{m5} +h_{C4}h_{m4} h_{C5}h_{m5} \right)
\end{equation}
\end{enumerate}
conveniently depicted in Fig \ref{fig:cascadedHandover}. A joint
optimisation of $\gamma$ and the integrator gains (both for the two
single integrators in the double-stage and the one for the instrument
controller) has not been conducted. Results may therefore be different
from the ones depicted in Fig \ref{fig:cascadedHandover}; however in a
cascaded handover the double-stage bandwidth will always limit the
overall bandwidth of the controller.

\section{Tilt anisoplanatism in laser-assisted AO tomography}

The NGS modes in laser-tomography AO are defined as the null modes
of the high-order LGS measurement space, i.e., modes that produce
average slope $\neq$ 0, but that due to the LGS tilt indetermination,
cannot be measured by the latter. In other words, the null space can
be thought of as the combination of all the modes that have non-null
projection onto the angle-of-arrival ( = not just Zernike tip and tilt
but also higher order Zernike modes). 



\subsection{Tilt tomography}

To estimate the tilt in the science direction of interest we consider the following options
\begin{enumerate}
\item Tilt tomography with spatio-angular reconstruction
  \cite{correia15, correia14}

Since for LTAO only pupil-plane tilt is required (no fitting on
multiple DMs) we use a simplified measurement model involving the pupil-plane turbulence only
\begin{equation}\label{eq:fwd_MOAO_meas_model}
\svec_\alphavec(t) = \D \phivec_\alphavec (t) + \etavec(t)
\end{equation}
The  \textit{minimum mean square error} (MMSE) tilt estimate assuming
$\svec$ and $\phivec$ are zero-mean and jointly Gaussian is
seamlessly found to be, for the $N_\beta$-science directions of interest \cite{andersonmoore_optimalfiltering05} 
\begin{align}\label{eq:MMSE_theor_MOAO}
\mathcal{E}\{\phivec_\betavec|\svec_\alphavec\}  \triangleq 
\CovMat_{(\phivec_\betavec,\svec_\alphavec)}\CovMat^{-1}_{\svec_\alphavec} {\svec}_\alphavec = \widehat{\phivec}_\betavec 
\end{align}
where $\mathcal{E}\{X|Y\} $ stands for mathematical expectation of $X$
conditioned to $Y$.  Since in general $\betavec \neq \alphavec$,
Eq. (\ref{eq:MMSE_theor_MOAO}) follows from
$\mathcal{E}\{\phivec_\betavec|\svec_\alphavec\} =
\mathcal{E}\{\phivec_\betavec|
\mathcal{E}\{\phivec_\alphavec|\svec_\alphavec\} \}$.  Matrices $\CovMat_{(\phivec_\betavec,\svec_\alphavec)}$ and
$\CovMat_{\svec_\alphavec}$ are spatio-angular covariance matrices that relate the tilt on direction
$\alphavec$ to that on direction $\betavec$. These matrices are computed
using formulae in \cite {whiteley98a} integrated in the simulator
OOMAO \cite {conan14}, provided knowledge of the $C_n^2$ atmospheric
profile is passed as input.
Figure \ref{fig:35layerMedianCn2Profile} depicts the 35-layer $C_n^2$
profile from ESO's site testing campaigns with a seeing of $0.65''$
(for a $r_0=0.151\,m@0.5\mu m$) and $L_0=25\,m$.
\begin{figure}[htpb]
	\begin{center}
            \includegraphics[width=0.58\textwidth]{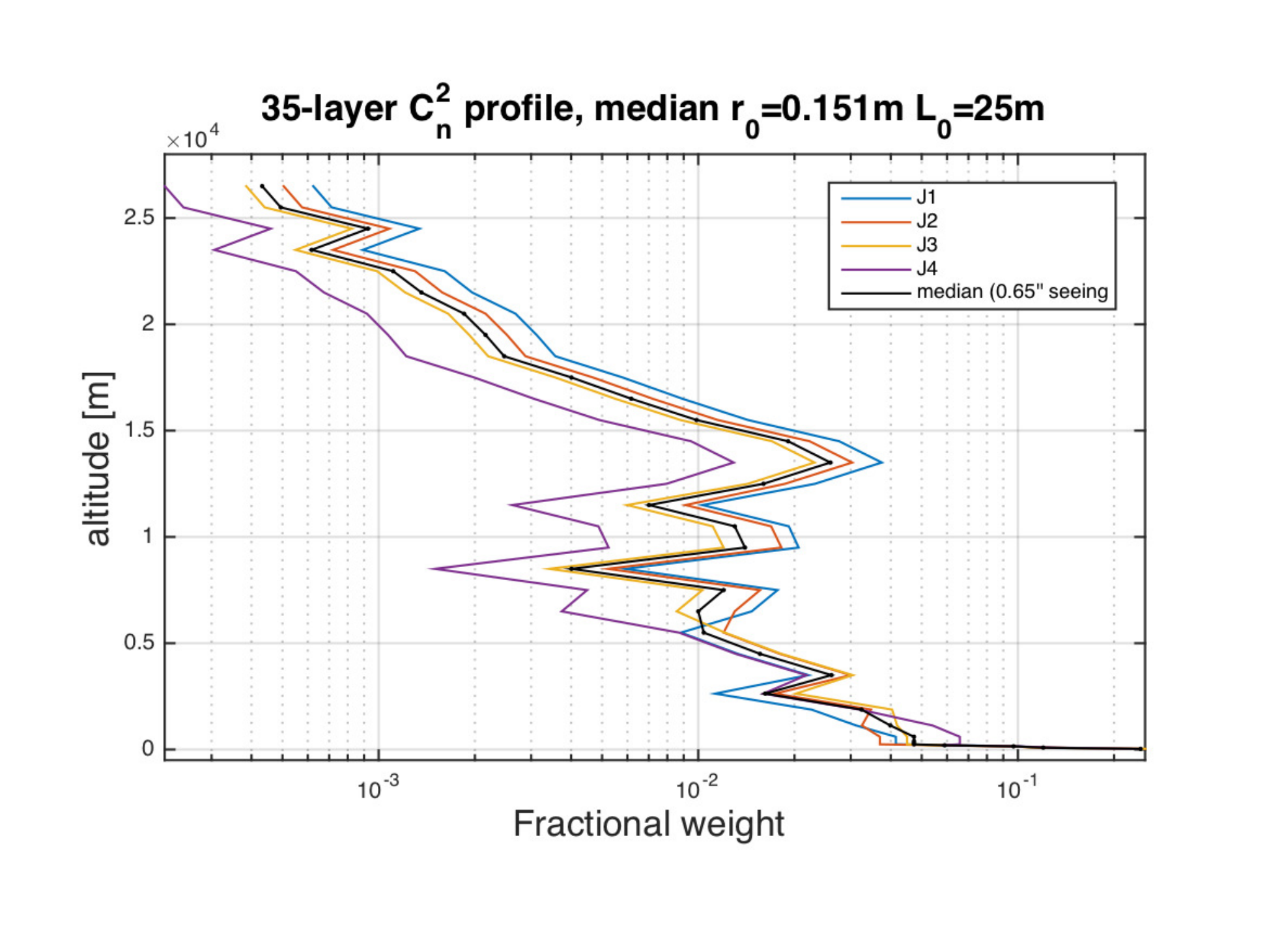}
	\end{center}
	\caption[]
	{\label{fig:35layerMedianCn2Profile}
      35-layer median $C_n^2$ profile;}
\end{figure}

Under the spatio-angular approach, it is also possible to compute
$\phivec(\rhovec, \betavec, t+\Delta)$, i.e., temporally predict the TT
ahead in time by adjusting the angles over which the correlations are
computed \cite {correia15}.
\begin{align}
\Emv = 
\CovMat_{(\phivec_\betavec,\svec_\alphavec)}\CovMat^{-1}_{\svec_\alphavec} {\svec}_\alphavec 
\end{align}
\item Isoplanatic tilt correction  (GLAO-like)
which consists in averaging the tilt measurements obtained
across the field 
\begin{align}
\Emv = \frac{1}{nGs}\sum_{i=1}^{nGs} \D^\dag{\svec}_{\alphavec,i} 
\end{align}
\item Tilt tomography using virtual DMs over two layers \cite{correia13}
\begin{align}
\Emv = \Proj_\betavec\left(\Proj_\alphavec^\T\D^\T\CovMat_\eta^{-1} \D \Proj_\alphavec\right)^{-1}\Proj_\alphavec^\T\D^\T\CovMat_\eta^{-1}
\end{align}
where we used the measurement model
\begin{equation}\label{eq:fwd_MOAO_meas_model}
\svec_\alphavec(t) = \D \Proj_\alphavec\varphivec (t) + \etavec(t)
\end{equation}
where $\D$ is an aperture-plane phase-to-gradient matrix representing
the SH-WFS and $\Proj$ a ray-tracing operator from atmospheric layers
to the pupil-plane.
 
This reconstructor is a noise-weighted reconstructor. In case of
equally noisy measurements it boils down to the simple averaging
(GLAO-like) case. 
\end{enumerate}

\subsection{Tomographic error in open-loop}

For the MMSE case, the standard performance assessment formulae
apply. 

Starting from
\begin{equation}\label{eq:res_var}
\sigma^2(\betavec) = \average{\left\Vert \phivec(\betavec) - \widehat{\phivec}(\betavec)\right\Vert^2}
\end{equation}
with $\widehat{\phivec}(\betavec) = \Rec  {\svec}(\alphavec)$ we get

\begin{align}\label{eq:res_var_developed}
\sigma^2(\betavec) & = \tr\{\CovMat_\betavec\} +
\tr\{\Rec\D\CovMat_\alphavec\D^\T\Rec^\T\}  \\ & -
\tr\{\Rec\D\CovMat_{\betavec,\alphavec}\}  -
\tr\{\CovMat_{\alphavec, \betavec}\D^\T\Rec^\T\}  \\ & +
                                                       \tr\{\Rec\CovMat_\etavec\Rec^\T\}
  \\
& = \sigma^2_{aniso} + \sigma^2_{noise}
\end{align}

For stars of equal magnitude, the virtual DMs tomography boils down to
the GLAO-like reconstructor as it is straightforward to show for diagonal noise
covariance matrices of equal entries. Simulations have shown that in
open loop the spatio-angular reconstructor is superior specially for
large off-axis distances. In what follows we will always use it and
leave a more complete comparison for later work.



\begin{figure}[htpb]
	\begin{center}
            \includegraphics[width=1.0\textwidth]{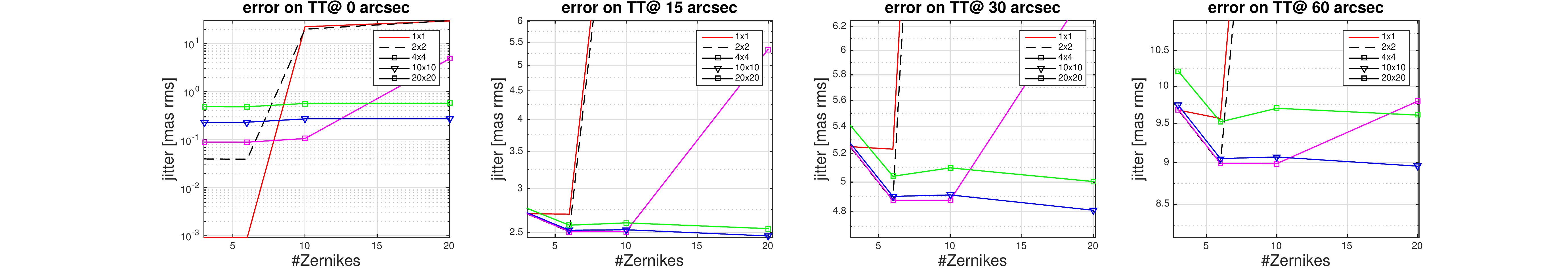}
	\end{center}
	\caption[]
	{\label{fig:err_fcn_nSubap_fcn_nZern}
      Tomographic error as a function of the number of Zernike modes
      estimated from measurements}
\end{figure}

Figure \ref{fig:err_fcn_nSubap_fcn_nZern} depicts the tomographic error as a function of the number of Zernike modes
estimated from measurements out of LOWFS from 1x1 to 20x20
sub-apertures. As expected, for on-axis observations the least
number of sub-apertures is preferred whereas for off-axis the
optimum seems to be a 10x10 sub-aperture SHWFS with 20+ Zernike
modes estimated. We understand this as a trade-off between the noise
propagation (more sub-apertures means more noise) and the pure
tomographic error (exploiting correlations of tilt with higher-order
modes is beneficial). 

\subsection{Error budget in closed-loop}
We splat the the errors as tomographic, noise and temporal under the
assumption that they are independent \cite{meimon10}
\begin{equation}
\sigma^2(\betavec) = \sigma^2_{aniso} + \sigma^2_{noise} + \sigma^2_{tempo}
\end{equation}
where $\sigma^2_{aniso} $ is computed as in Eq. (\pageref{eq:res_var})
and
\begin{equation}
\sigma^2_{noise} =  \tr\{\Rec\CovMat_{\etavec'}\Rec^\T\}
\end{equation}
with the horizontal elements of the loop-filtered noise covariance
matrix populated with
\begin{equation}
\sigma^2_{\etavec'} = 2 T_s \int_0^{1/2/T_s} NTF(\nu) \dint\nu \,\,\, \sigma^2_{\etavec}
\end{equation}
with $NTF(\nu)$ from Eq. (\ref{eq:NTF}).
The temporal error
\begin{equation}
\sigma^2_{tempo} = \int_0^{1/2/T_s} RTF(\nu)  PSD_\varphi(\nu)\dint\nu 
\end{equation}
where $RTF(\nu)$ is gathered for different controller options from
Eq. (\ref{eq:RTF}).

\section{Assessing sky-coverage}

In this section we investigate the residual jitter as a function of
the field and the photon flux.  We have used the LQG+AR2 shown
previously which provided the least isoplanatic tilt residual -- see
Table \ref{tab:comparisonControllers}.
Figure \pageref{fig:jitter} shows the residual jitter with one or two
TT measurements in the field (symmetric around the origin, same flux
on both). We obtain a 1.2\, mas\,rms jitter for $10^6$ PDE/aperture/$m^2$ on axis
with 1 SH and only a small improvement with 2 SH.

On the opposite end, residual jitter of roughly 5.5\,mas\,rms for 10
PDE/aperture/$m^2$ 1 arcmin off-axis which lowers to 5\,mas\,rms in
case 2 TT stars are used.
\begin{figure}[htpb]
	\begin{center}
            \includegraphics[width=0.45\textwidth]{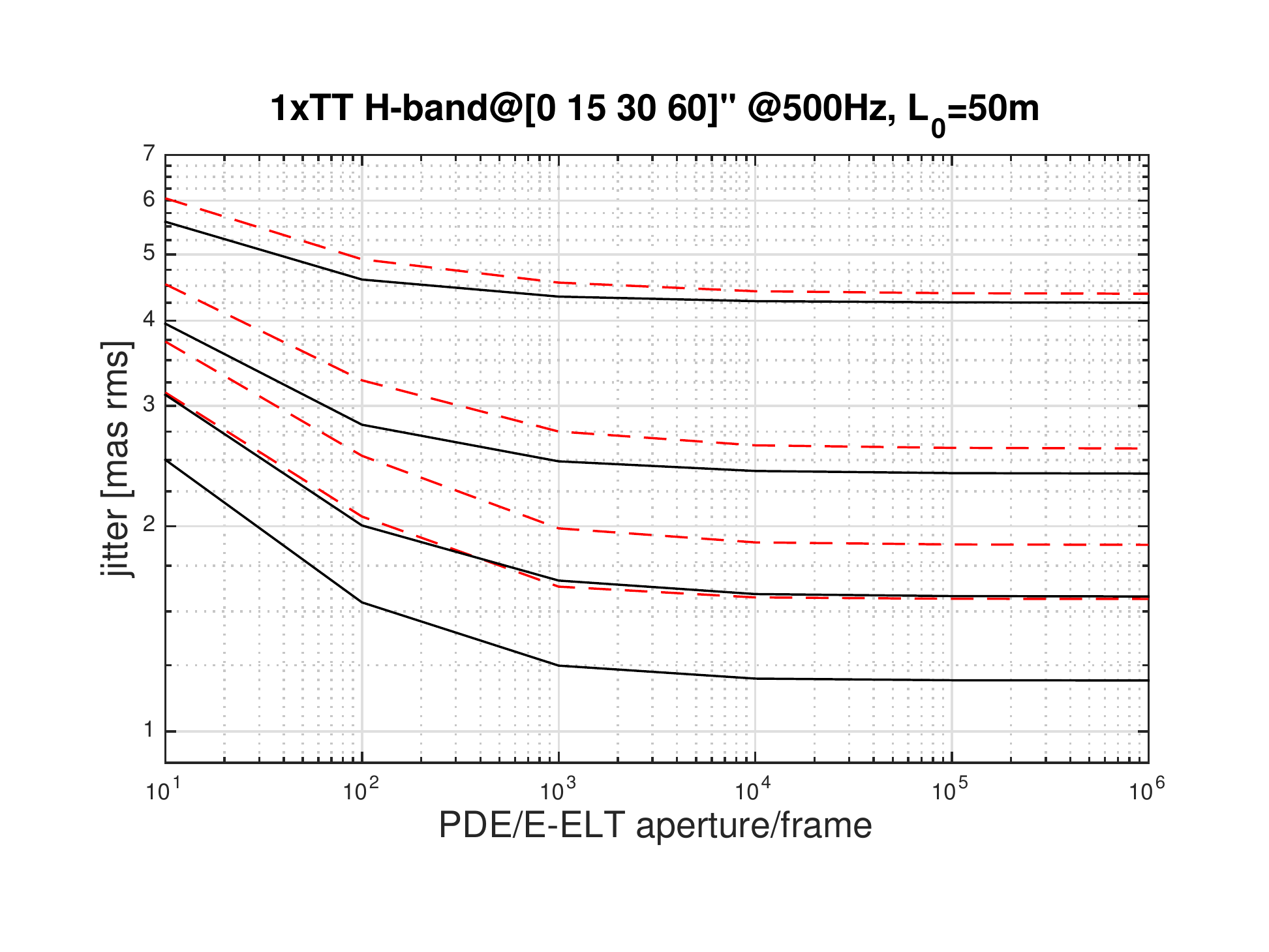}\includegraphics[width=0.45\textwidth]{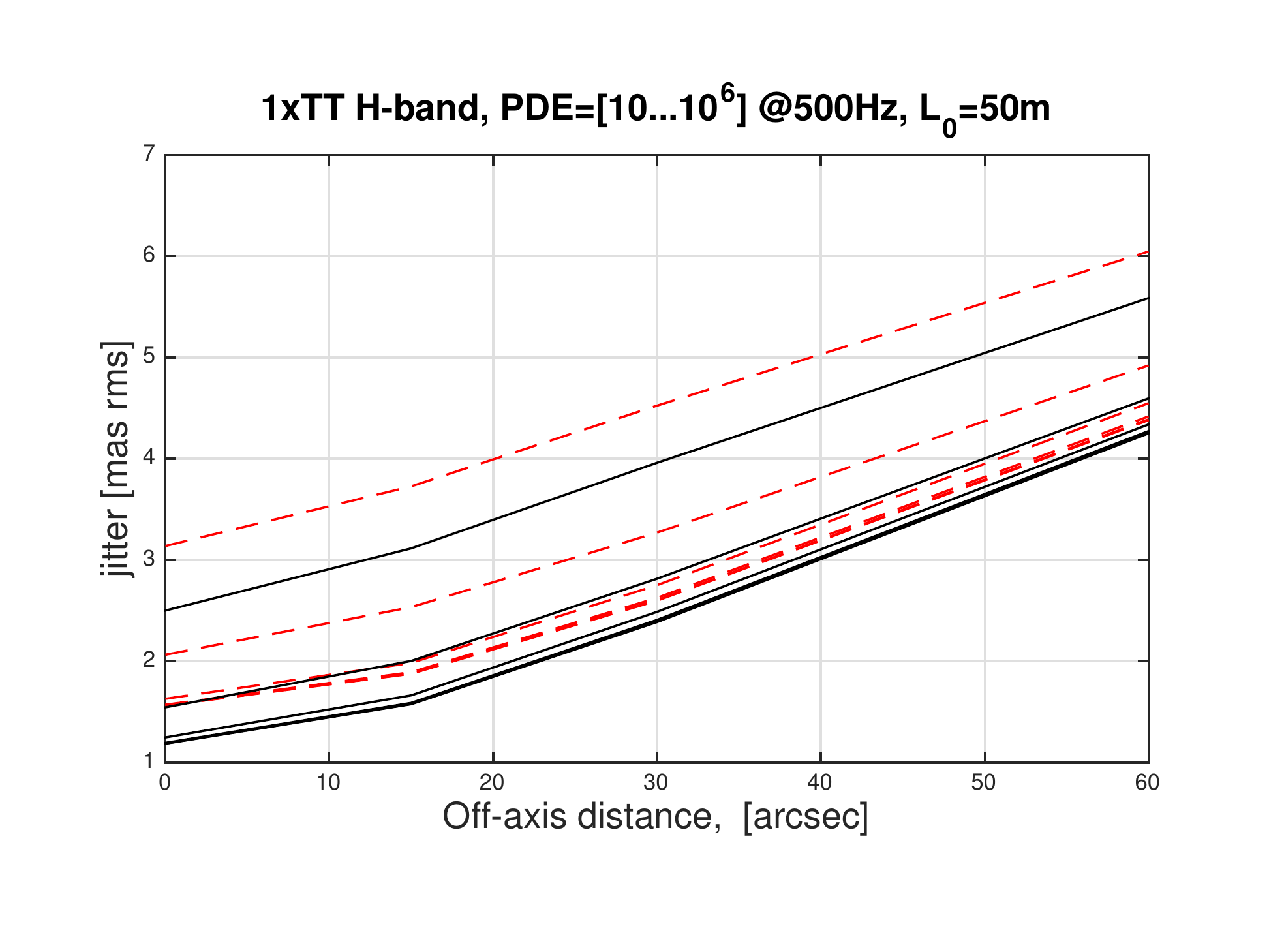}
 \includegraphics[width=0.45\textwidth]{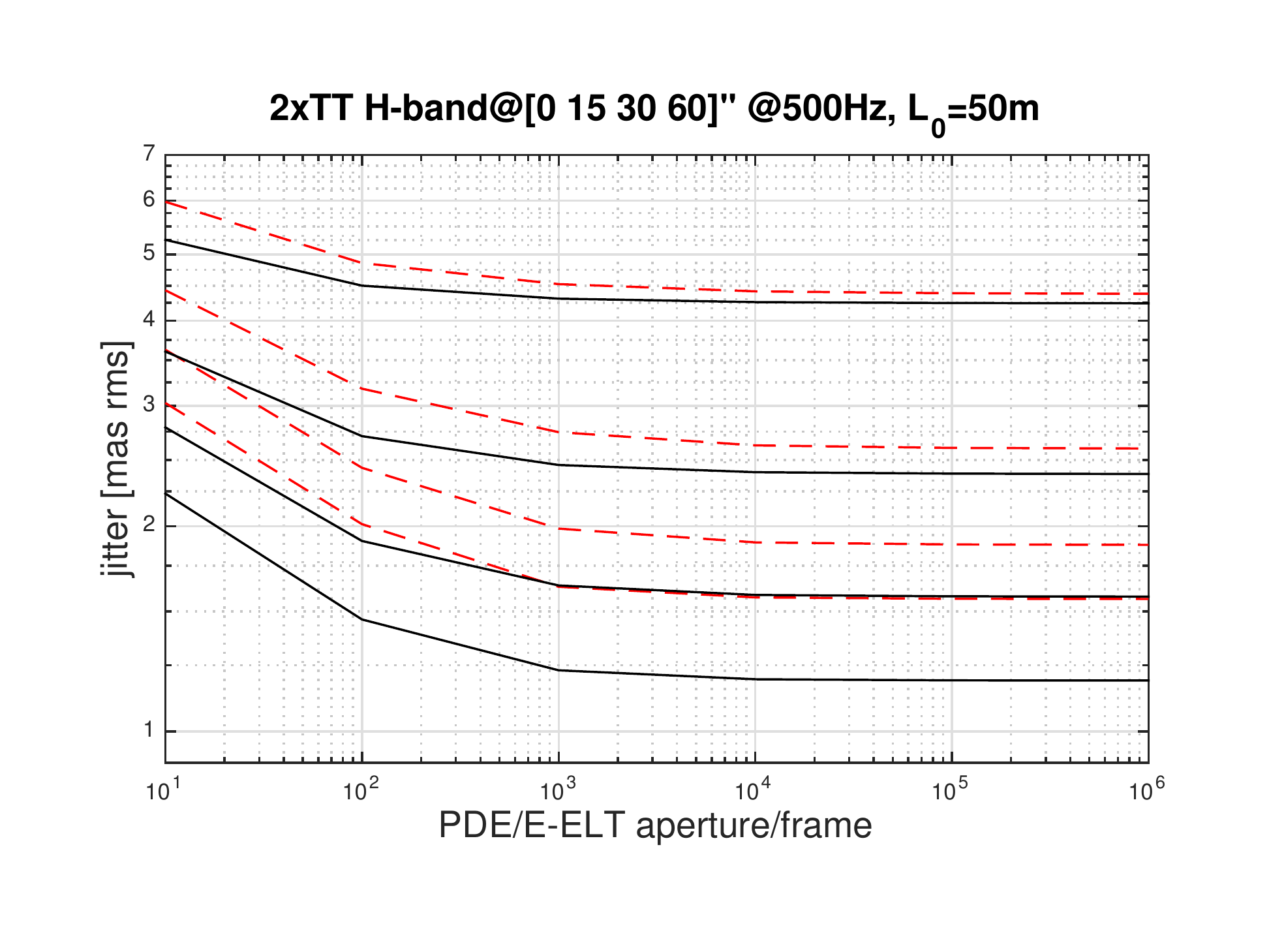}\includegraphics[width=0.45\textwidth]{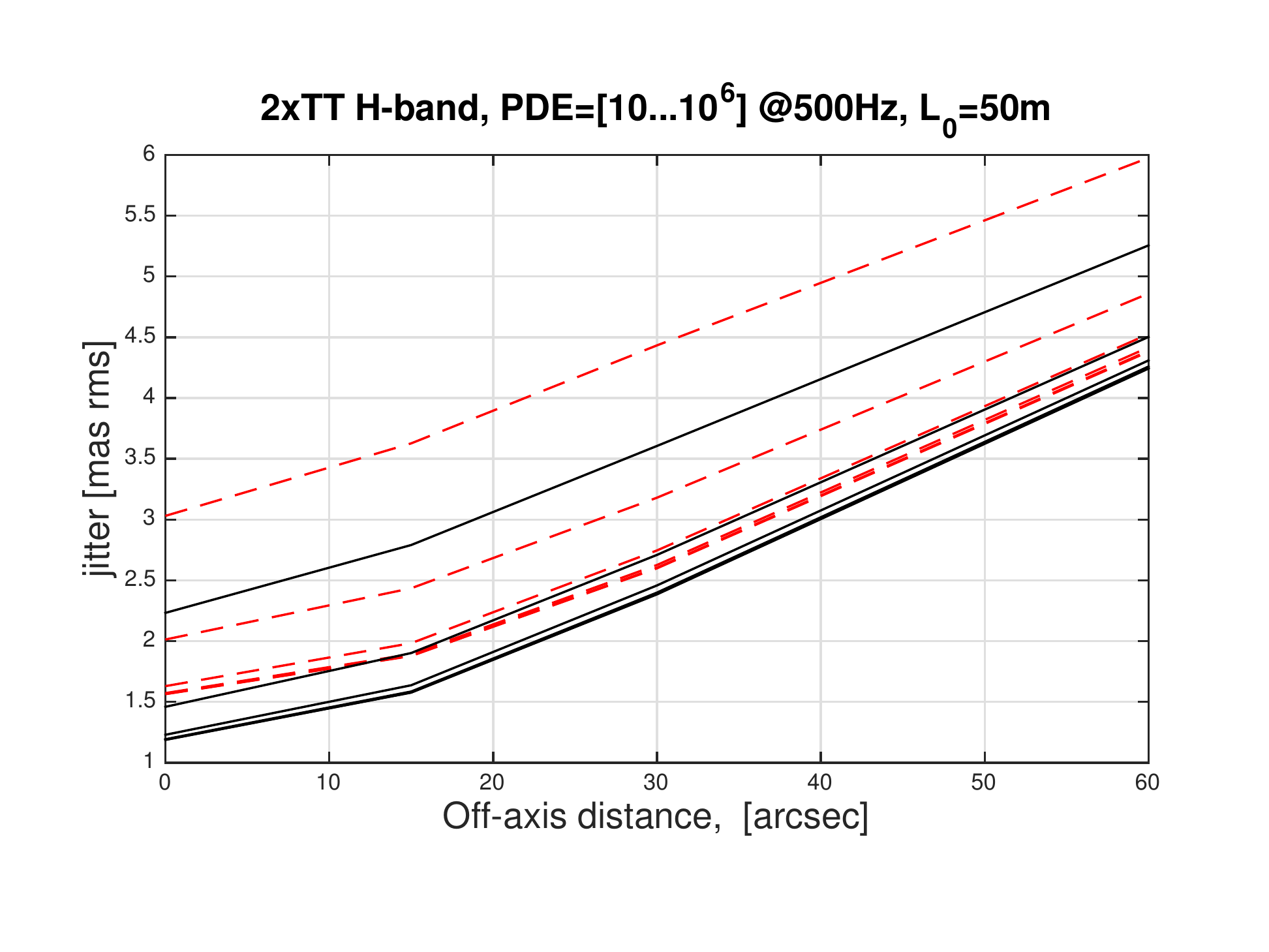}
	\end{center}
	\caption[]
	{\label{fig:jitter}
      (Top-panel) Left: Residual jitter (ATM+wind-shake on 1 axis) as a function
      of the PDE/aperture/frame using a single-aperture SH-WFS probing
    a 35-layer ESO atmospheric profile with $L_0$= 50\,m at angles [0, 15,
    30, 60] arcsec. Red: optimised double integrator+lead filter,
    black: LQG. Left: same data plotted as a function of the
    off-axis-distance for different noise levels as a function of the
    PDE rangin from 10 to $10^6$. (Bottom-panel): same as top-panel with a second
    TT for secondary guiding.}
\end{figure}

\begin{figure}[htpb]
	\begin{center}
            \includegraphics[width=0.8\textwidth]{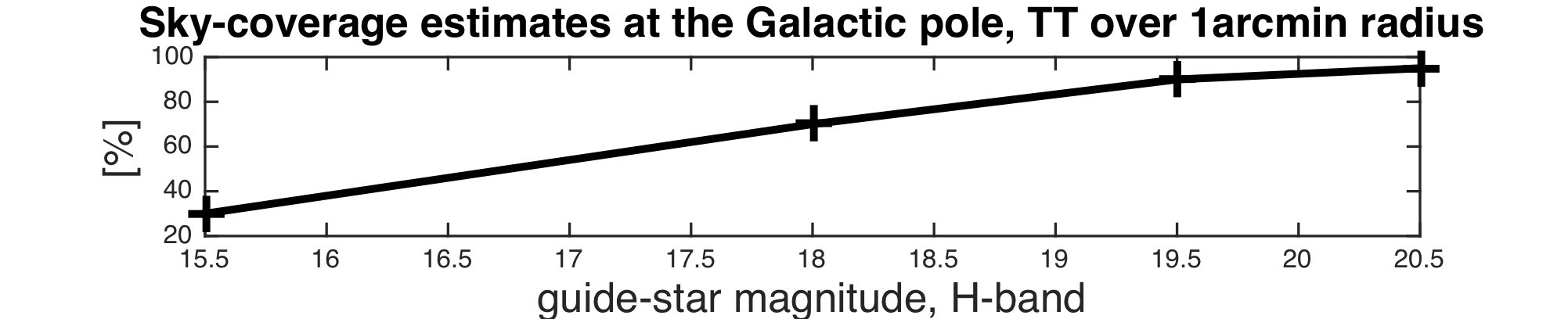}
	\end{center}
	\caption[]
	{\label{fig:skyCoverage}
     Sky-coverage estimates at the galactic pole. }
\end{figure}

 With these values we could compute sky-coverage estimates at the
 galactic pole considering a wide-band tilt measurement from I- to
 K-band in a field of 1\,arcmin radius -- as is shown in
 Fig. \ref{fig:skyCoverage} as a function of the H-band magnitude\cite{neichel16}.

\section{Outlook}
Our simulations of isoplanatic tilt control for the Harmoni project
show that LQG controllers in stand-alone mode (i.e. sequential
handover) ensuring a 40db/dec at low-frequencies with 500Hz temporal frame-rate
ensure the least residual whilst being stable even in the presence of
non-stationary transients affecting the system every 5 minutes. Work
on the cascaded handover mode where the instrument and telescope's TT
controllers work in tandem needs further parameter optimisation,
although it seems to us a less promising path at this time.

Regarding tilt anisoplanatism we've found
\begin{itemize}
\item the use of secondary guiding improves the use of 1 single tilt
  measurement in the field only slightly (less than 10\%)
\item use of single-aperture SH for tilt measurement is best in the
  centre of the field whereas a 10x10 SH seems a better option when
  using a start farther off-axis up to 1 arcmin radius
\item the spatio-angular tilt tomography controller ensures least
  residuals when compared to a GLAO-like reconstructor averaging the
  tilt in the field or to the virtual-DM reconstructor used in MCAO
\end{itemize}

Our preliminary sky-coverage estimates indicate 70\% for H- magnitude
guide-stars at the galactic pole over 1 arcmin radius.

\acknowledgments     

The research leading to these results received the support of the
A*MIDEX project (no. ANR-11-IDEX-0001-02) funded by the ”Investissements
d'Avenir” French Government program, managed by the French National
Research Agency (ANR).

All the simulations and analysis done with the object-oriented MALTAB AO
simulator (OOMAO) \cite {conan14} freely available from
https://github.com/cmcorreia/LAM-Public


\begin{thebibliography}{10}

\bibitem{correia13}
Correia, C., V\'{e}ran, J.-P., Herriot, G., Ellerbroek, B., Wang, L., and
  Gilles, L., ``Increased sky coverage with optimal correction of tilt and
  tilt-anisoplanatism modes in laser-guide-star multiconjugate adaptive
  optics,'' {\em J. Opt. Soc. Am. A}~{\bf 30},  604--615 (Apr 2013).

\bibitem{correia15}
Correia, C.~M., Jackson, K., V\'{e}ran, J.-P., Andersen, D., Lardi\`{e}re, O.,
  and Bradley, C., ``Spatio-angular minimum-variance tomographic controller for
  multi-object adaptive-optics systems,'' {\em Appl. Opt.}~{\bf 54},
  5281--5290 (Jun 2015).

\bibitem{thatte16}
Thatte, N.~A. and the Harmoni~consortium, ``The e-elt first light spectrograph
  harmoni: capabilities and modes,'' in [{\em Proc. of the
  SPIE}{\nolinebreak\hspace{0.1em}]},  {\em Ground-based and Airborne
  Instrumentation for Astronomy}~{\bf 9908},  9908--71, SPIE (2016).

\bibitem{neichel16}
Neichel, B. and the Harmoni~consortium, ``The harmoni laser tomography
  module,'' in [{\em Proc. of the SPIE}{\nolinebreak\hspace{0.1em}]},  {\em
  Adaptive Optics Systems}~{\bf 9909},  9909--1, SPIE (2016).

\bibitem{sauvage16}
Sauvage, J.-F. and the Harmoni~consortium, ``Status of the harmoni single
  conjugate adaptive optics module,'' in [{\em Proc. of the
  SPIE}{\nolinebreak\hspace{0.1em}]},  {\em Adaptive Optics Systems}~{\bf
  9909},  9909--82, SPIE (2016).

\bibitem{roddier99}
Roddier, F.,  [{\em Adaptive Optics in Astronomy}{\nolinebreak\hspace{0.1em}]},
  Cambridge University Press, New York (1999).

\bibitem{sedghi16}
Babak~Sedghi, Michael~Müller, M.~D., ``Analysing the impact of vibrations on
  e-elt primary segmented mirror,'' in [{\em Proc. of the
  SPIE}{\nolinebreak\hspace{0.1em}]},  {\em Modeling, Systems Engineering, and
  Project Management for Astronomy}~{\bf 9911},  9911--39, SPIE (2016).

\bibitem{correia12}
Correia, C., V\'{e}ran, J.-P., and Herriot, G., ``Advanced vibration
  suppression algorithms in adaptive optics systems,'' {\em J. Opt. Soc. Am.
  A}~{\bf 29},  185--194 (Mar 2012).

\bibitem{correia12b}
Correia, C., V{\'e}ran, J.-P., Herriot, G., Ellerbroek, B., Wang, L., and
  Gilles, L., ``Advanced control of low order modes in laser guide star
  multi-conjugate adaptive optics systems,'' in [{\em Proc. of the
  SPIE}{\nolinebreak\hspace{0.1em}]},   84471S--84471S--12 (2012).

\bibitem{correia10a}
Correia, C., Raynaud, H.-F., Kulcs{\'a}r, C., and Conan, J.-M., ``On the
  optimal reconstruction and control of adaptive optical systems with mirror
  dynamics,'' {\em J. Opt. Soc. Am. A}~{\bf 27},  333--349 (Feb. 2010).

\bibitem{andersonmoore_optimalcontrolLQG05}
Anderson, B. D.~O. and Moore, J.~B.,  [{\em Optimal Control, Linear Quadratic
  Methods}{\nolinebreak\hspace{0.1em}]}, Dover Publications Inc. (1995).

\bibitem{neichel09}
Neichel, B., Fusco, T., and Conan, J.-M., ``Tomographic reconstruction for
  wide-field adaptive optics systems: Fourier domain analysis and fundamental
  limitations,'' {\em J. Opt. Soc. Am. A}~{\bf 26}(1),  219--235 (2009).

\bibitem{raynaud16}
Henri-François G. Raynaud Remy~Juvenal, Caroline~Kulcsar, C.~P., ``The
  control switching adapter: a practical way to ensure bumpless switching
  between controllers while ao loop is engaged,'' in [{\em Proc. of the
  SPIE}{\nolinebreak\hspace{0.1em}]},  {\em Adaptive Optics Systems}~{\bf
  9909},  9909--193, SPIE (2016).

\bibitem{correia12a}
Correia, C. and {V{\'e}ran}, J., ``Woofer-tweeter temporal correction split in
  atmospheric adaptive optics,'' {\em Opt. Lett.}~{\bf 37} (Aug 2012).

\bibitem{correia11e}
Correia, C., ``Gpi tt controller,'' Tech. Rep.~1, Herzberg Institute of
  Astrophysics (2011).

\bibitem{correia14}
Correia, C., Jackson, K., V\'{e}ran, J.-P., Andersen, D., Lardi\`{e}re, O., and
  Bradley, C., ``Static and predictive tomographic reconstruction for
  wide-field multi-object adaptive optics systems,'' {\em J. Opt. Soc. Am.
  A}~{\bf 31},  101--113 (Jan 2014).

\bibitem{andersonmoore_optimalfiltering05}
Anderson, B. D.~O. and Moore, J.~B.,  [{\em Optimal
  Filtering}{\nolinebreak\hspace{0.1em}]}, Dover Publications Inc. (1995).

\bibitem{whiteley98a}
Whiteley, M.~R., Welsh, B.~M., and Roggemann, M.~C., ``Optimal modal wave-front
  compensation for anisoplanatism in adaptive optics,'' {\em J. Opt. Soc. Am.
  A}~{\bf 15},  2097--2106 (1998).

\bibitem{conan14}
Conan, R. and Correia, C., ``Object-oriented matlab adaptive optics toolbox,''
  in [{\em Proc. of the SPIE}{\nolinebreak\hspace{0.1em}]},   {\bf 9148},
  91486C--91486C--17 (2014).

\bibitem{meimon10}
Meimon, S., Fusco, T., Clenet, Y., Conan, J.-M., Assémat, F., and Michau, V.,
  ``The hunt for 100\% sky coverage,'' (2010).

\end{thebibliography}

\end{document}